\documentclass[]{spie}  
\usepackage[]{graphicx}
\usepackage{geometry}
\title{A new era of wide-field submillimetre imaging: on-sky performance of SCUBA-2  } 

\author{ Jessica T. Dempsey\supit{a}, Wayne S. Holland\supit{b,c}, Antonio Chrysostomou\supit{a}, David S. Berry\supit{a}, Daniel Bintley\supit{a}, Edward L. Chapin\supit{c}, Simon C. Craig\supit{a}, Iain M. Coulson\supit{a}, Gary R. Davis\supit{a}, Per Friberg\supit{a}, Tim Jenness\supit{a}, Andy G. Gibb\supit{c}, Harriet A. L. Parsons\supit{a}, Douglas Scott\supit{c}, Holly S. Thomas\supit{a}, Remo P. J. Tilanus\supit{a,d}, Ian Robson\supit{b}, Craig A. Walther\supit{a}
\skiplinehalf
\supit{a}Joint Astronomy Centre, 660, N. A’ohoku Place, University Park, Hilo, HI 96720, United States\\
\supit{b}UK Astronomy Technology Centre, Royal Observatory, Blackford Hill, Edinburgh EH9 3HJ\\
\supit{c}Institute for Astronomy, University of Edinburgh, Royal Observatory, Blackford Hill, Edinburgh EH9 3HJ\\
\supit{d}Department of Physics \& Astronomy, University of British Columbia, 6224 Agricultural Road, Vancouver BC V6T 1Z1, Canada\\
\supit{e} Netherlands Organization for Scientific Research ,Laan van Nieuw Oost-Indie 300, NL2509 AC The Hague, The Netherlands\\
}

\authorinfo{Further author information: (Send correspondence to J.T.D.)\\J.T.D.: E-mail: j.dempsey@jach.hawaii.edu, Telephone: 1 808 969 6566\\}

\begin{document} 
  \maketitle 

\begin{abstract}
SCUBA-2 is the largest submillimetre wide-field bolometric camera ever built. This 43 square arc-minute field-of-view instrument operates at two wavelengths (850 and 450 microns) and has been installed on the James Clerk Maxwell Telescope on Mauna Kea, Hawaii. SCUBA-2 has been successfully commissioned and operational for general science since October 2011. This paper presents an overview of the on-sky performance of the instrument during and since commissioning in mid-2011. The on-sky noise characteristics and NEPs of the 450$\mu$m and 850$\mu$m arrays, with average yields of approximately 3400 bolometers at each wavelength, will be shown. The observing modes of the instrument and the on-sky calibration techniques are described. The culmination of these efforts has resulted in a scientifically powerful mapping camera with sensitivities that allow a square degree of sky to be mapped to 10 mJy/beam rms at 850$\mu$m in 2 hours and 60 mJy/beam rms at 450$\mu$m  in 5 hours in the best weather.\\

\end{abstract}



\section{INTRODUCTION}
\label{sec:intro} 

SCUBA-2 is a dual-wavelength camera containing 5120 pixels in each of two focal planes. Each focal plane consists of 4 separate rectangular sub-arrays each with 1280 bolometers butted together to give the full field of approximately 8 $\times$ 8 arcmin.  The two wavebands, which target the atmospheric windows centred at 450$\mu$m and 850$\mu$m (as was its predecessor, SCUBA~\cite{holland1999}),  observe the sky simultaneously by means of a dichroic beamsplitter. SCUBA-2 was delivered from the UK Astronomy Technology Centre to the Joint Astronomy Centre in Hawaii in April 2008 with one engineering sub-array at each waveband. The focal-planes were fully-populated with science-grade arrays in summer 2010 and the first astronomical data with the full complement of arrays was taken in early-2011. 

Submillimetre observations are more critical than ever before in a broad range of astronomical areas of interest. SCUBA-2 is presently the largest submillimetre camera in the world, and the fastest wide-area ground-based submillimetre imager. An ambitious set of legacy surveys is currently well underway and these are designed to maximise the impact of SCUBA-2's wide-field, fast-mapping capabilities and to complement the current and future set of submillimetre facilities, such as Herschel and ALMA. The surveys include large-scale surveys of  molecular clouds and star-formation regions (Gould-Belt Survey and Galactic Plane Survey), an extensive survey of nearby galaxies (NGLS), targeted study of debris discs around nearby stars (SONS), a deep cosmology survey (S2CLS) and a broad survey of the outer Galaxy (SaSSy). 

This paper will outline the performance of SCUBA-2 as it undertakes its first semester of full science observations, following the successful commissioning of the instrument in October 2011. The instrument characterization results will be presented, including its noise performance on the sky, responsivity and bolometer yields. The observing strategies will be discussed, and the final instrument sensitivities will be shown. Finally, a selection of the first science observations will be presented. A detailed article on the instrument performance is given in Ref.~\citenum{holland2012}.\\

\section{On-sky Commissioning} 

Two science-grade sub-arrays were delivered for early commissioning in 2009. The remaining six science-grade sub-arrays were delivered to Hawaii and installed in the instrument in mid-2010. Previous characterisation results are presented in Ref.~\citenum{bintley2010}. The full account of the instrument technical performance is given in Ref.~\citenum{bintley2012} in these proceedings. The SCUBA-2 bolometers are integrated arrays of superconducting transition-edge sensors (TESs) with a transition temperature T$_c$.  The devices are voltage-biased to a temperature near the centre of its transition range, whereupon any temperature fluctuation results in a current change through the TES. This induces a magnetic field in the superconducting quantum interference device (SQUID) bump-bonded to each TES, and this signal is then amplified using chains of SQUIDs after which the amplified signal is digitised. SCUBA-2 detectors have individually-coupled resistive heaters which are used to compensate for changes in incident optical power, thereby allowing the detector bias point to remain constant over a wide-range of sky powers. The initial commissioning focused on optimising the array performance in the dark, ie. with the shutter closed. Once the dark performance had been determined, the on-sky commissioning tasks commenced as follows:

\begin{itemize}
  \item On-sky array setup: confirming that the dark setup parameters (including TES and heater bias values, and SQUID flux offsets and biases) produced similar performance with the shutter open. 
  \item Flat-fielding: determining the optimal parameters to measure the bolometer responsivity.
  \item Optical power loading: equalize the calibration of the sub arrays by updating the heater coupling values using the relative optical response of the arrays.
  \item Noise performance: assessment of the noise equivalent power (NEP) of the arrays at each wavelength, and full investigation of the noise properties of the arrays, including short- and long-term stability. 
  \item Atmospheric calibration: determining the two filter extinction coefficients to correct for waveband-dependent atmospheric attenuation.
  \item Observing modes: optimising scan-speeds and patterns to produce the most efficient maps over the largest areas, while determining limiting factors.
  \item Data reduction: refining the iterative map-maker as the noise properties of the arrays and characteristics of the data were assessed.
  \item Flux calibration: absolute conversion from the calibrated picowatt output of the detectors to janskys, using observations of sources of known submillimetre flux.
  \item Sensitivity: after completion of the above tasks, determination of the full-system sensitivity.
\end{itemize}

\subsection{Flat-fielding}\label{flat}
 \begin{figure}
   \begin{center}
   \begin{tabular}{c}
   \includegraphics[height=6cm]{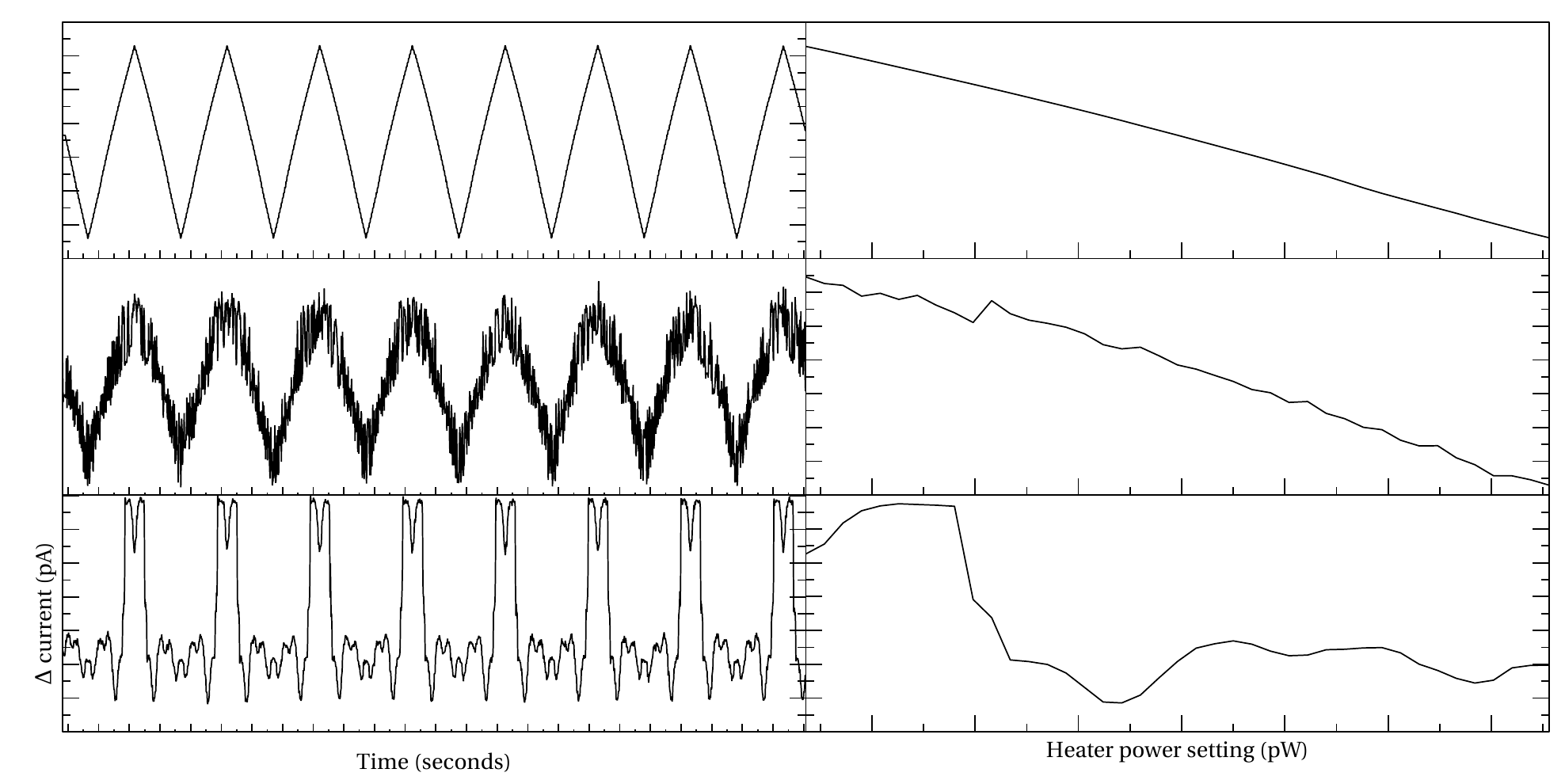}
   \end{tabular}
   \end{center}
   \caption{Examples of the $\sim$3 pW peak-to-peak fast ramps of the heater power that are used to calculate the flat-field response (right) of each individual TES bolometer. The heater is ramped eight times in a ten-second interval. The top plots show a typical, well-behaved bolometer ramp and the corresponding current change as a function of heater power, with the responsivity determined from a fit to the inverted slope $dI/dP$. Bolometers are flagged and removed as bad if showing too poor a S/N (middle plots) or a severe non-linear response (bottom). \label{fig:flatfield} 
}
   \end{figure} 
 
 On-sky observations with SCUBA-2 proceed following an array setup, conducted with the shutter closed, to optimise the SQUID settings.  When the shutter is opened, the power from the resistive heater coupled to each bolometer is decreased to compensate for the incident optical power. The resistors coupling to the TES varies between sub-arrays. The ``effective'' resistance can be calculated by measurement of the I-V curve obtained at different heater power settings. These resistances were shown to be significantly different for each individual sub-array. Additionally, the relative optical response of the arrays was determined by measurement of the signal when observing bright sources such as Jupiter. The coupling factors were adjusted such that the picowatt response to the planet signal was equal in all sub-arrays. The I-V curves were used to obtain the average absolute pW calibration of the bolometers and then the flux-determined factors were used to calibrate the relative sub-array response, as this is ultimately the measure of the optical response through the entire system from an astronomical source.\\

 The heaters are then employed to calibrate the individual bolometer responsivity. The response of the detectors is determined by measuring the slope of the current/power change as the heater current below each pixel is ramped quickly over a small range (a few pW peak-to-peak), as shown in the left-hand plots in Figure~\ref{fig:flatfield}. The flat-field response is the inverse gradient of the current change as a function of heater power (or $dI/dP$) of each individual bolometer. At this point the quality of the bolometer performance is assessed: bolometers with responsivities above or below a threshold limit are removed, along with bolometers whose response, $dI/dP$, is non-linear, or shows too poor a S/N. Figure~\ref{fig:flatfield} shows a good bolometer trace (top) and example of a bolometer that was flagged resulting from a poor S/N (centre) and extreme non-linear response (bottom).

Initially, this flat-fielding was completed in the dark with the shutter closed, assuming that the heater-track as the shutter open would compensate sufficiently for the change in optical power. Testing showed, however, that better bolometer yields and more stable operation was achieved if this flat-fielding was conducted again once the shutter was open, particularly when there was sky instability.

\subsection{Noise equivalent power}\label{nep}

 \begin{figure}
   \begin{center}
   \begin{tabular}{c}
   \includegraphics[height=7cm]{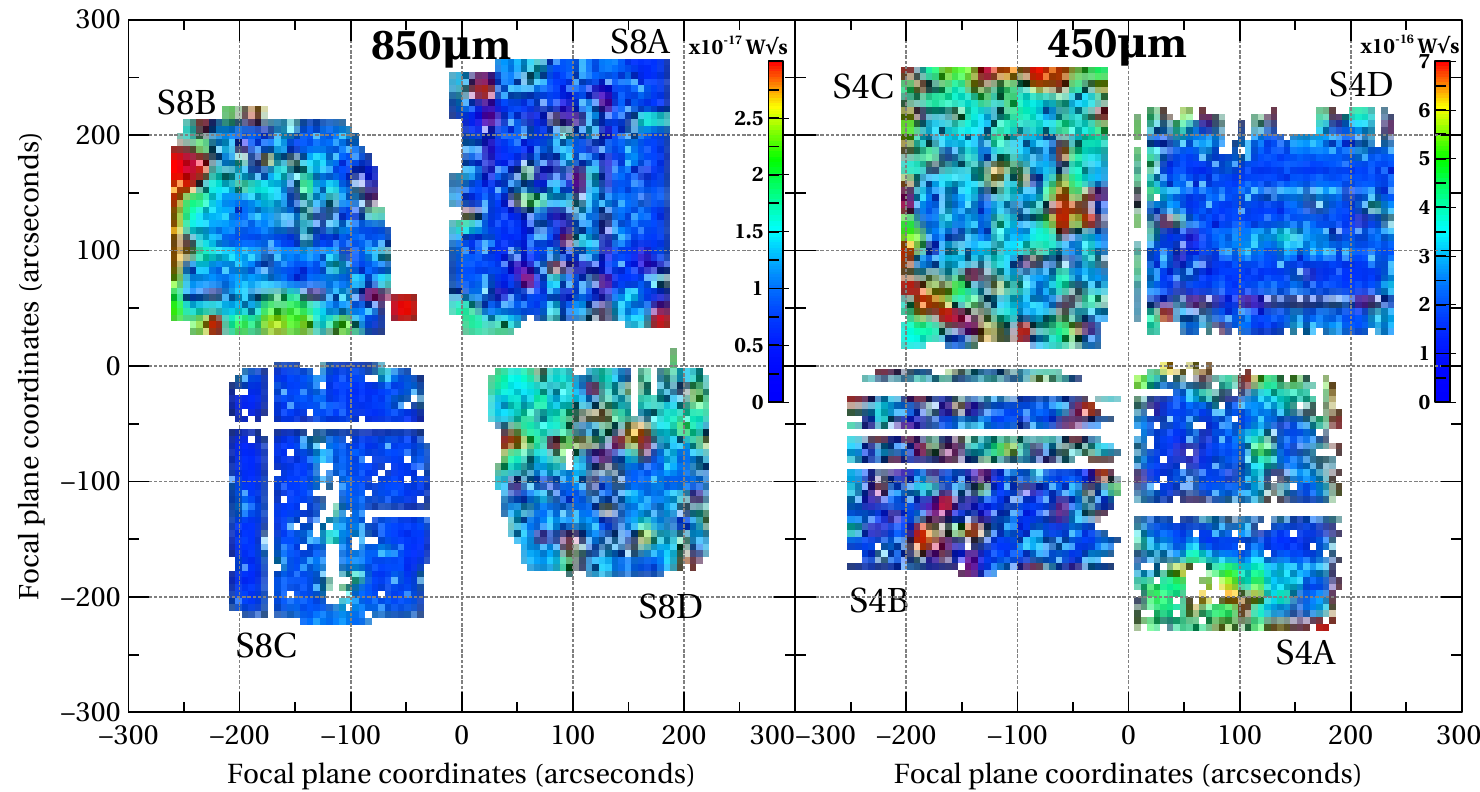}
   \end{tabular}
   \end{center}
   \caption{Typical NEP image of the 850$\mu$m focal-plane on the sky (left) and the 450$\mu$m NEP focal plane on the sky (right). The colour scale is indicated by the bar on the right-hand side and is in units of $\times 10^{-17}$/$\times 10^{-16}$ W$\sqrt{s}$ for the 850/450$\mu$m respectively. \label{fig:nepboth} 
}
   \end{figure}

 \begin{figure}
   \begin{center}
   \begin{tabular}{c}
   \includegraphics[width=14cm]{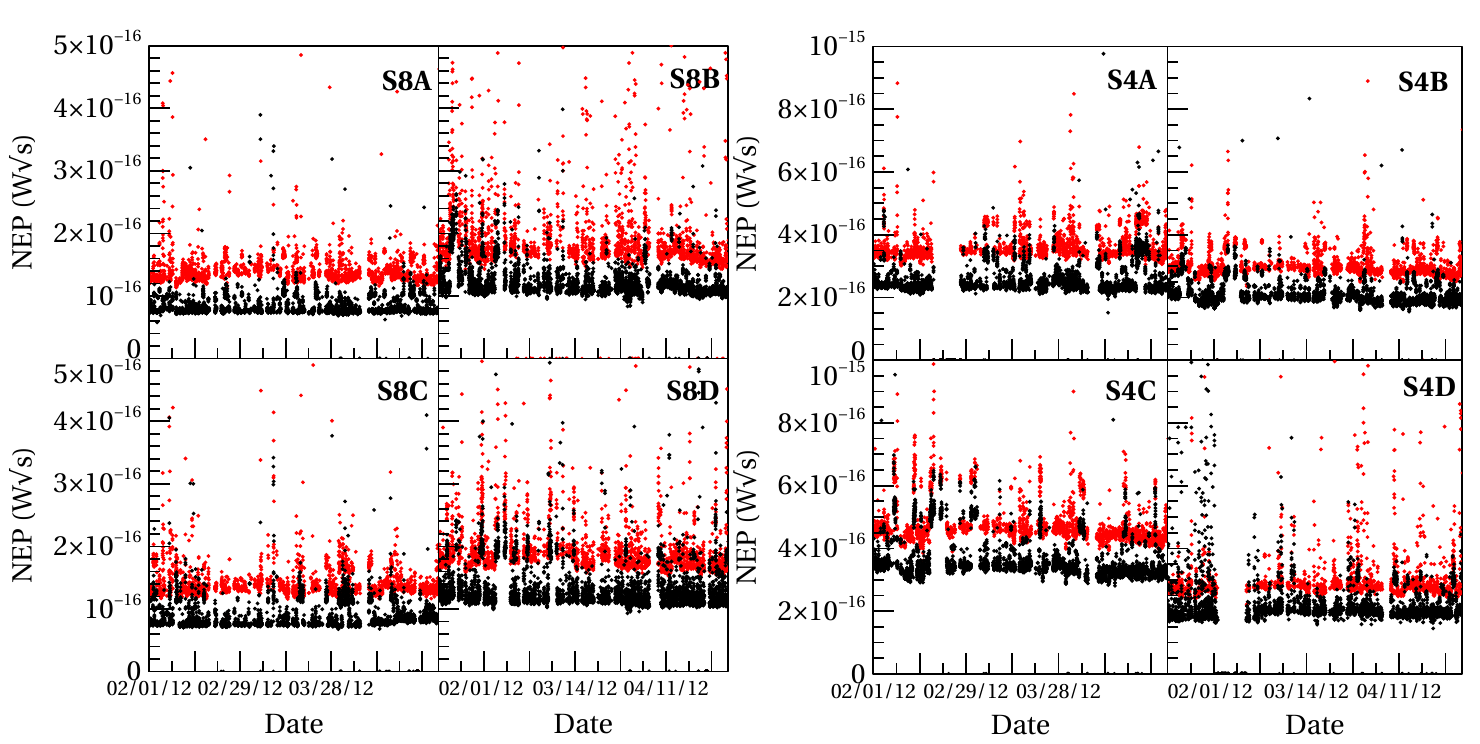}
   \end{tabular}
   \end{center}
   \caption{A short, five-second dark frame is taken at the start of every observation. The weighted-mean NEP of each sub-array is plotted versus time for each dark noise measurement (black points) taken between February and May 2012.  The weighted-mean NEP for the first five-seconds of the corresponding sky observation is plotted in red. The upwards scatter in the NEPs is caused by variations in incident sky power.\label{fig:neptime} 
}
   \end{figure}

The responsivity is calculated by measuring the change in output current of the TES as a function of heater power, producing a quantity in units of A/W. The NEP (Noise equivalent power) is given in units of W$\sqrt{s}$ and in practice is proportional to the rms of the noise in one second divided by the responsivity. Figure~\ref{fig:nepboth} shows a snapshot of the 450$\mu$m (top) and 850$\mu$m focal plane NEPs for a typical observation.\\

\begin{figure}
   \begin{center}
   \begin{tabular}{c}
   \includegraphics[height=7cm,width=14cm]{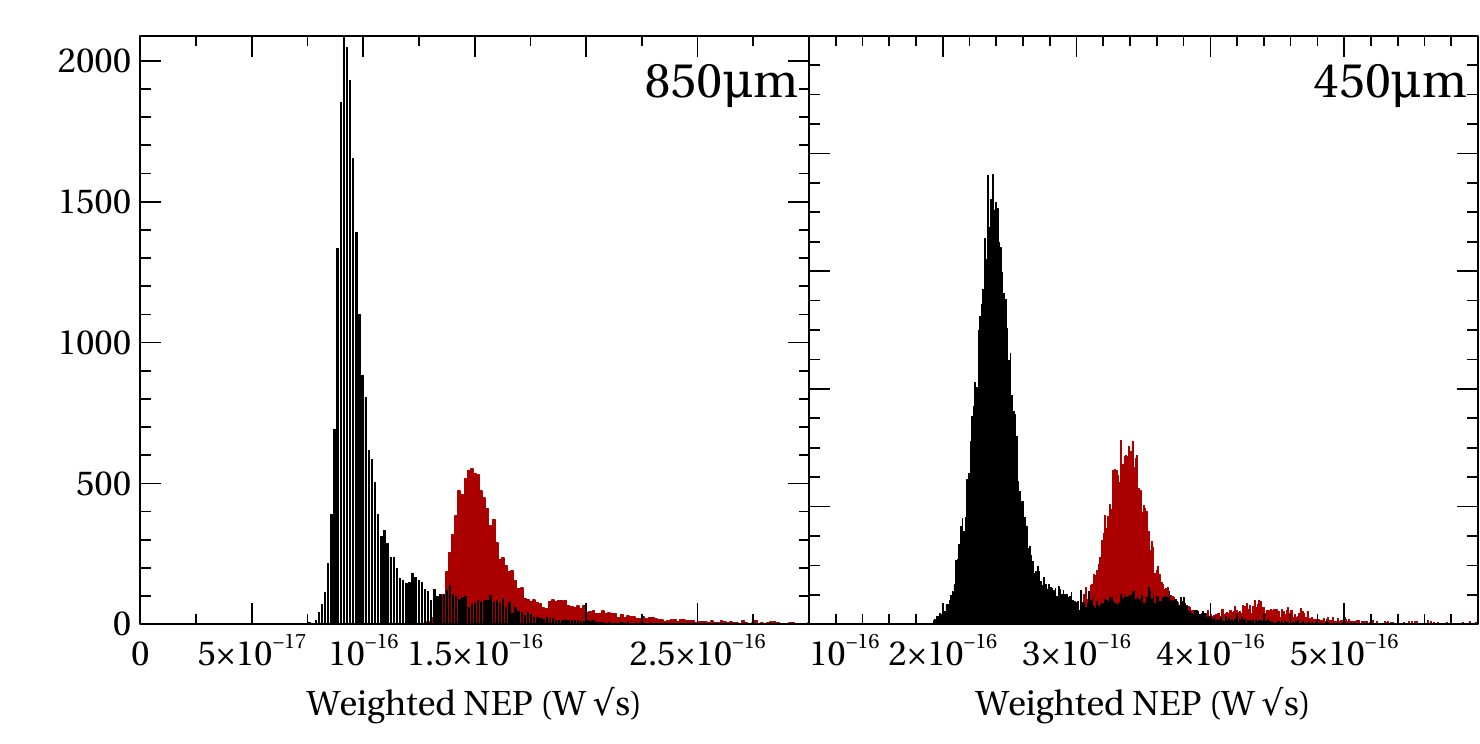}
   \end{tabular}
   \end{center}
   \caption{The histogram distribution of the weighted-mean NEPs from Figure~\ref{fig:neptime}, averaged over all sub-arrays. The dark NEP histograms are in black at both wavelengths and the sky NEP histograms are red. The sky NEPs will be affected by variations in sky power, so it is expected that this distribution should be wider. However what this clearly shows is the small secondary distribution of a higher noise state, both in the dark and on the sky, most clearly seen at 450$\mu$m. This increased noise is observed on nearly every sub-array at some point every few days (though is rarely correlated between them), and investigation into the cause is ongoing.\label{fig:nephist} 
}
   \end{figure}

Figure~\ref{fig:neptime} shows the weighted mean NEP for each sub-array in the dark (black points) and on the sky (red points) at both wavelengths, for every observation taken between February and May 2012. For the majority of the time the dark NEP performance shows good stability. The sky NEP is a function of the atmospheric conditions at the time of the observation, thus exhibiting more upwards scatter. The distribution of the mean NEP for the full-array at each wavelength over this time period is shown in Figure~\ref{fig:nephist}. What can be noted in both the time series and the histogram distribution is a distinct mode of higher noise performance, particularly noticeable at 450$\mu$m. The noise increase occurs randomly, perhaps once every few nights, and seems uncorrelated between sub-arrays, as well as wavelength bands. The noise performance seems unaffected by setting up the arrays again, and often returns to its lower, normal noise performance without any clearly correlated process or event. This, and other noise characteristics, are still under investigation and are discussed further in Ref.~\citenum{holland2012} and \citenum{bintley2012}.\\

\begin{table}[h]
\caption{Average array performance.}
\label{table1}
\begin{center}
\begin{tabular}{|c|c|c|c|} 
\hline
\rule[-1ex]{0pt}{3.5ex}  Wavelength & Average yield (dark/sky) & Mean NEP (dark/sky)& Mean responsivity (dark/sky)\\
\rule[-1ex]{0pt}{3.5ex}  [$\mu$m] &  & [W$\sqrt{s}$]& [A/W]  \\
\hline
\rule[-1ex]{0pt}{3.5ex}   850 & 3430/3339  &  $9.3 \times 10^{-17}$/$1.48\times 10^{-16}$  & $1.4\times 10^{6}$/$1.4\times 10^{6}$  \\
\hline
\rule[-1ex]{0pt}{3.5ex}  450  & 3540/3432  &  $2.36\times 10^{-16}$/$3.4\times 10^{-16}$    & $6.1\times 10^{5}$/$6.1\times 10^{5}$ \\
\hline
\end{tabular}
\end{center}
\end{table}

 The average on-sky array performance statistics are presented in Table~\ref{table1} for the data shown in Figure~\ref{fig:neptime}. If the heater power adequately compensates for the optical loading as the shutter is opened, and the absolute calibration of the power is correct, the responsivities should be identical on sky as when measured in the dark, and this is confirmed in the results shown.\\

\section{Calibration}

To convert detector power into astronomical flux units, corrections must be applied to account for (a) atmospheric extinction (mostly a result of absorption by atmospheric water vapour); and (b) the transmission properties of the optics, which account for the effective area of the detector, the spectral response of the bolometers (bandpass), and additional optical efficiency factors – which is termed the flux conversion factor (FCF). A full account of the SCUBA-2 calibration can be found in Ref.~\citenum{dempsey2012}.\\

\subsection{Atmospheric attenuation}\label{atm}
    \begin{figure}
   \begin{center}
   \begin{tabular}{c}
   \includegraphics[height=8cm,width=12cm]{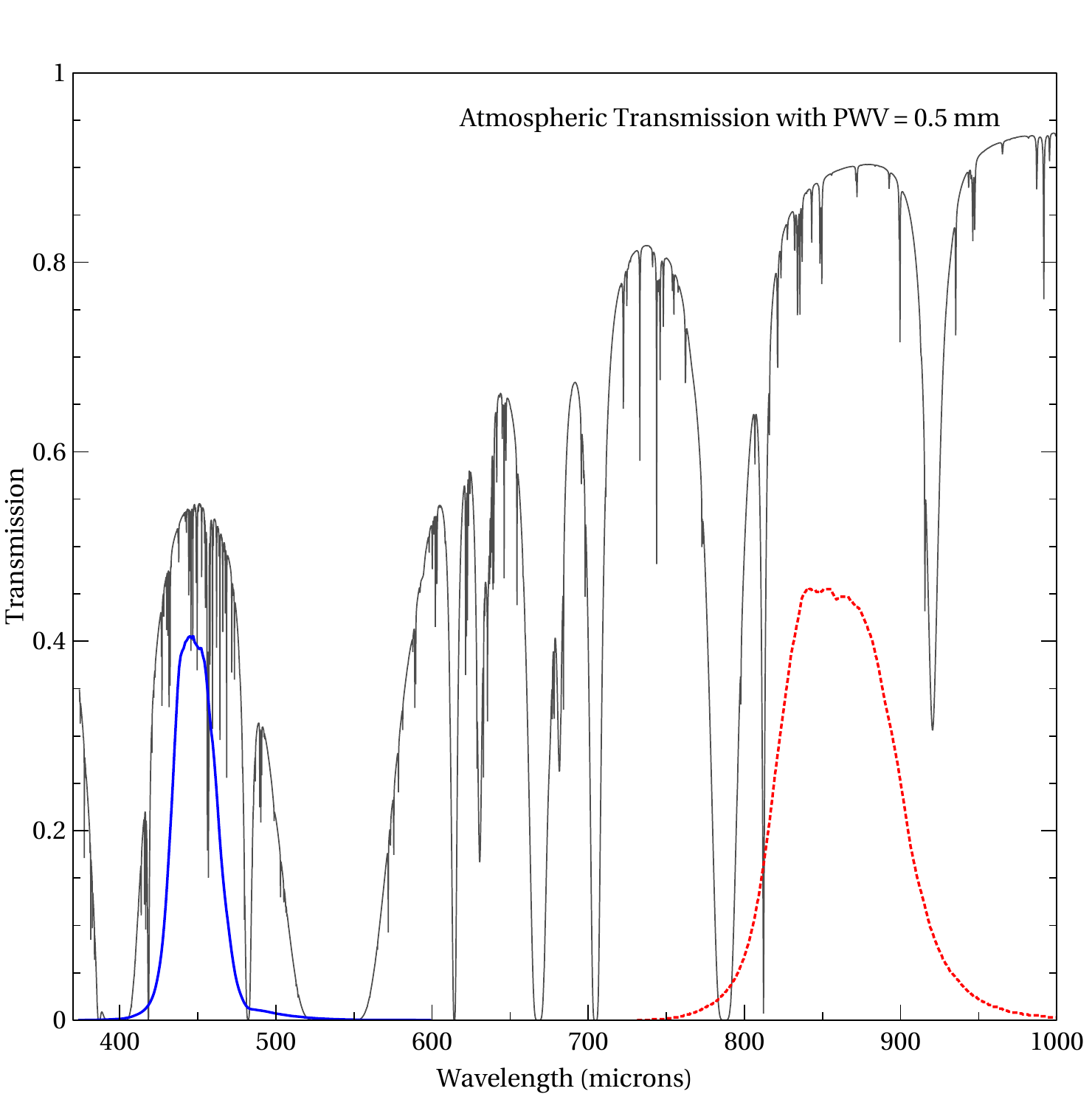}
   \end{tabular}
   \end{center}
   \caption{450$\mu$m  (red solid) and 850$\mu$m (blue dotted) bandpass filters superimposed on the submillimetre atmospheric transmission curve for Mauna Kea assuming 0.5 mm of precipitable water vapour.\label{fig:atmtrans} 
}
   \end{figure}

The atmosphere limits ground-based observations at submillimetre wavelengths, providing only a few semi-transparent windows even at a high, dry site such at the summit of Mauna Kea, Hawaii. The SCUBA-2 bandpasses at both wavelengths are shown superimposed on the submillimetre atmospheric transmission windows in Figure~\ref{fig:atmtrans}. The extinction coefficient, $\tau$,  must be explicitly determined for each observing waveband.  The wavelength-dependent extinction relations for SCUBA were calculated as a function of the CSO tipper 225$\,$GHz opacity~\cite{archibald}, measured every fifteen minutes and which has operated throughout the lifetime of both SCUBA and SCUBA-2. The JCMT has a 183$\,$GHz water vapour monitor (WVM) installed in the receiver cabin which measures the precipitable water vapour (PWV) along the line of sight at 1.2 second intervals~\cite{wiedner}. The WVM has the advantage of better time resolution, which is important given that SCUBA-2 collects data at a rate of 200$\,$Hz. Also, by measuring directly along the line-of-sight of the telescope we are not limited to the fixed azimuth of the CSO tipper, and we do not need to assume a plane-parallel atmosphere.\\

The extinction coefficients along the line-of-sight, at each SCUBA-2 wavelength, will be described in terms of the PWV determined by the water vapour monitor. The relation was derived by collating the calibration observations of sources of known flux taken over the six month on-sky commissioning period from May 2011 until October 2011, and then during operations until May 2012. These observations were reduced without applying any assumed extinction correction and then fitted using a least-squares algorithm, where the precipitable water vapour (PWV) is measured in millimetres. The resulting extinction coefficients at each wavelength are:
\begin{equation}
\tau_{850} =  0.179 \cdot ({\mbox{PWV}} + 0.337)
\end{equation}
\begin{equation}
\tau_{450} =  1.014 \cdot ({\mbox{PWV}} + 0.142)
\end{equation}

For reference, the SCUBA-2 extinction correction coefficients in terms of the opacity at 225$\,$GHz are: $\tau_{\mbox{\tiny 850}} =  4.6 \cdot (\tau_{\mbox{\tiny 225}} - 0.0043)$ and $\tau_{\mbox{\tiny 450}} =  26.0 \cdot (\tau_{\mbox{\tiny 225}} - 0.012)$, to allow easy comparison with the previous extinction terms for SCUBA.

\subsection{Flux calibration} 

    \begin{figure}
   \begin{center}
   \begin{tabular}{c}
   \includegraphics[height=10cm,width=12cm]{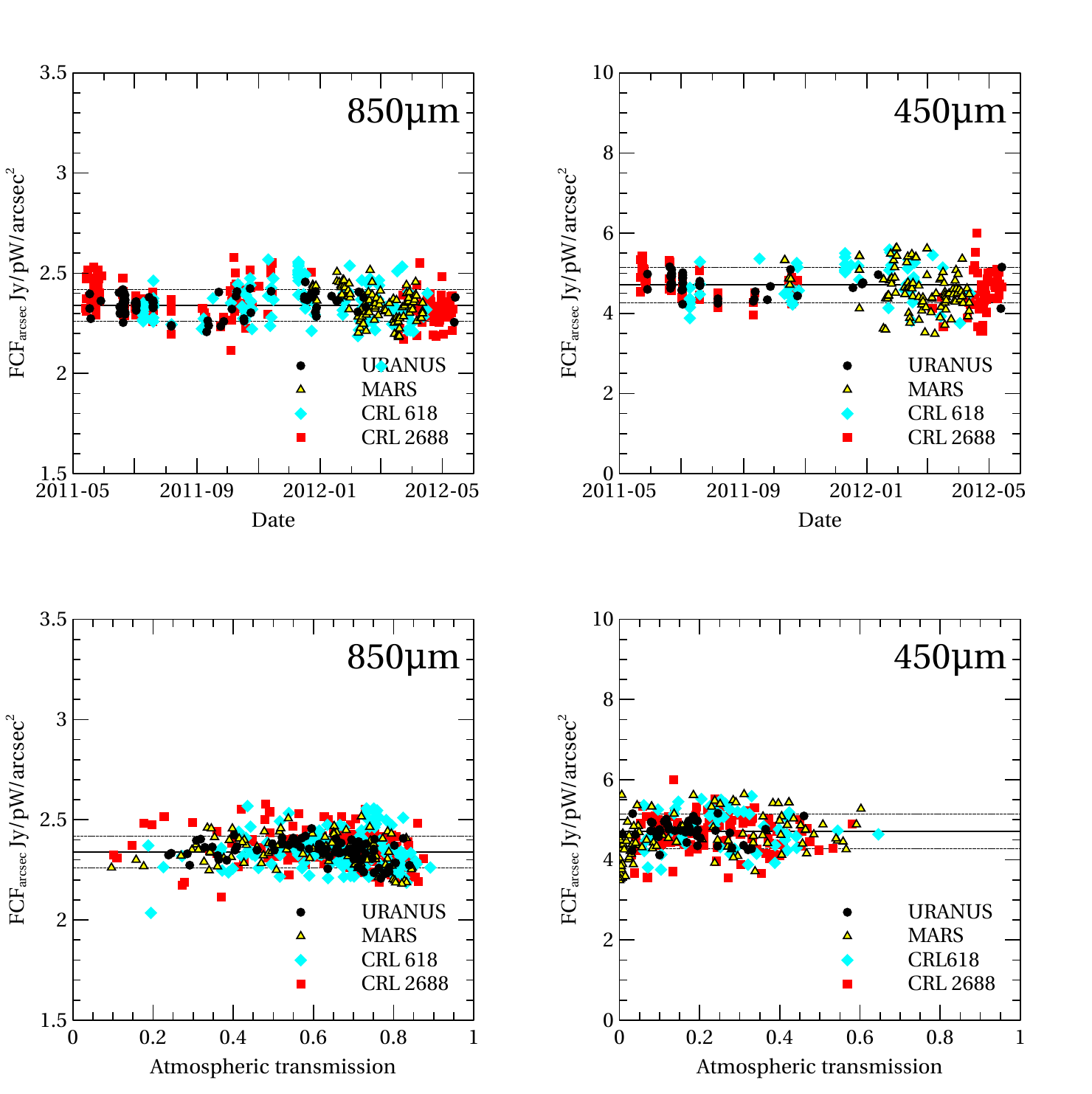}
   \end{tabular}
   \end{center}
   \caption{Flux conversion factor, FCF$_{\mbox {\tiny arcsec}}$ results, at 850$\mu$m (left) and 450$\mu$m (right), plotted as a function of time (top) between May 2011 and May 2012, and as a function of atmospheric transmission (bottom).  \label{fig:fcfasec} 
}
   \end{figure}

  \begin{figure}
   \begin{center}
     \begin{tabular}{c}
         \includegraphics[height=8cm]{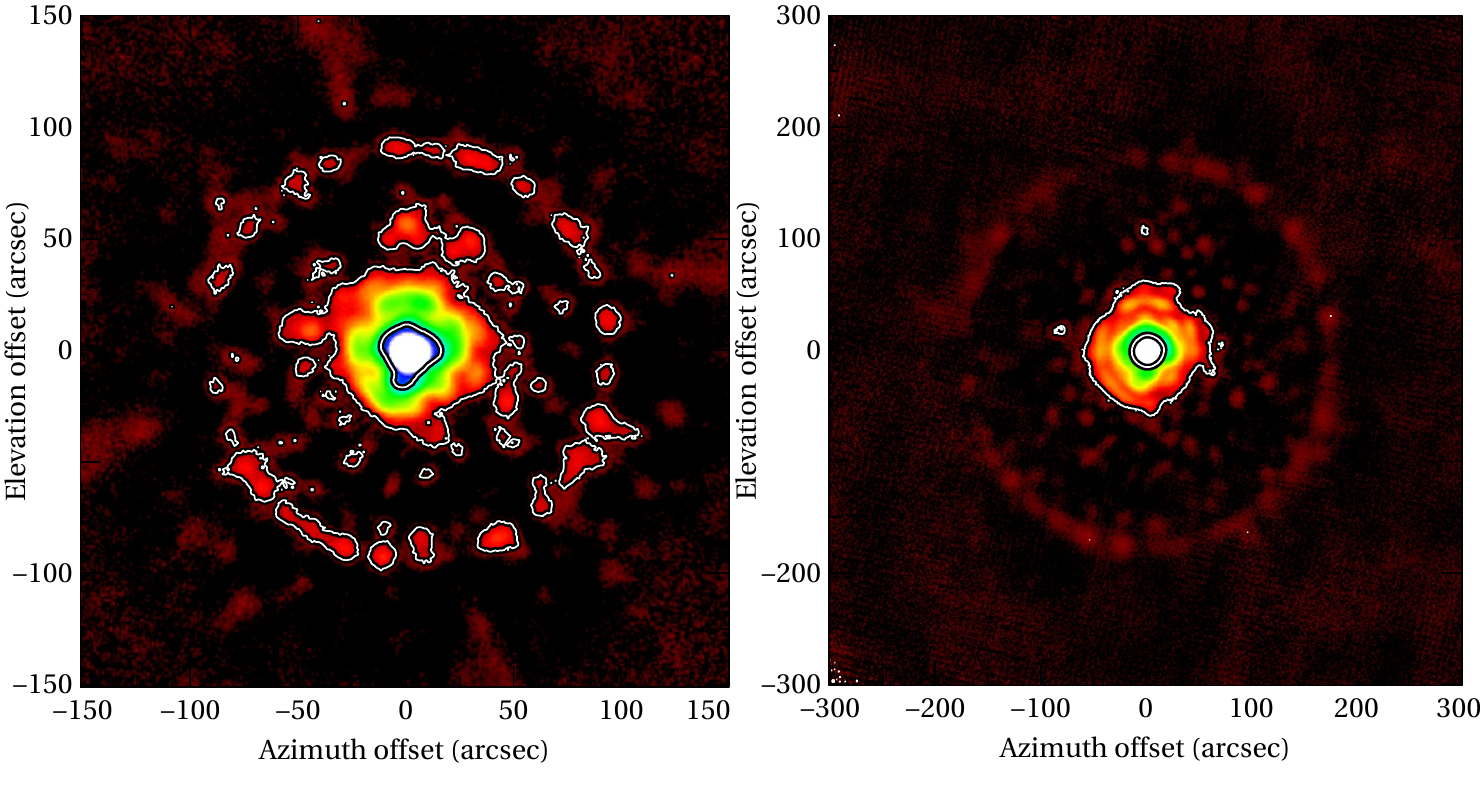} 
     \end{tabular}
   \end{center}
   \caption{Left: A 5$\times$5 arcmin image of a typical 450$\mu$m MARS map, with contours at 5\% and 0.1\% peak flux. The colour scale is a square-root scaling of the emission below 10\%. Right:A 10$\times$10 arcmin image of a typical 850$\mu$m Mars map, with contours at 5\% and 0.1\% of the peak flux. The colour scale is a square-root scaling of the emission below 10\%. \label{fig:beam450} 
}
   \end{figure} 

Flux calibration is achieved by observation of astronomical sources with known flux properties. To convert the measured signal of a source from picowatts to janskys a flux conversion factor (FCF) must be applied to the map. If the optical performance is stable, and flat-fielding of the instrument and extinction correction are properly calculated, the FCF should be a constant factor.The FCF is simply described as the ratio between the known flux $S_{\mbox {\tiny known}}$ in Jy and the measured flux $S_{\mbox {\tiny meas}}$, the signal in pW summed over the source. To apply the FCF when calculating a flux integrated over a source, the size of the pixels in the output map must also be taken into account. This factor is called the FCF$_{\mbox {\tiny arcsec}}$, and is given by:
\begin{equation}
{\mbox {FCF}}_{\mbox {\tiny arcsec}} = \frac{S_{\mbox {\tiny known}}}{S_{\mbox {\tiny meas}} \times {\mbox {pixscale}}^{2}}
\label{fcfeq1}
\end{equation}

\noindent {where the pixel scale is in arcseconds, giving the units of  Jy/pW/arcsec$^{2}$. If wishing to measure the peak flux of a source, the peak FCF is simply the ratio of the known peak flux in Jy to the measured peak flux in pW.}

Uranus and Mars were the primary calibrators observed during the observing period presented here. Uranus and Mars brightness temperatures are obtained using the JCMT FLUXES software, in which the Uranus brightness temperature is taken from Ref.~\citenum{wright} and the Mars results are derived from a model by Ref.~\citenum{moreno}. Secondary calibrators observed on a nightly basis were CRL618 and CRL2688. The calibrations were taken between May 2011 and May 2012, and over 500 observations (at each wavelength) were included in this analysis. The total integrated flux for each calibrator map was calculated in a 60$\,$arcsec aperture, with the background level calculated in an annulus with inner radius 90$\,$arcsec and outer radius 120$\,$arcsec (1.5 and 2.0 times the source aperture). Figure~\ref{fig:fcfasec} show the results of the FCF calculations for the standard calibrators, Uranus, Mars, CRL618 and CRL2688. In the left shows the $FCF_{\mbox {\tiny arcsec}}$ as a function of time (top), and versus atmospheric transmission(bottom). At both wavelengths the results show no systematic deviation, indicating the instrument and optical performance was consistent during this year-long period.\\

Figure~\ref{fig:beam450} show a typical map of Mars, at both wavelengths, colour-scaled between 0 and 10\% of the peak emission to highlight the low-level structure in the map which is produced by the telescope optics. Two contours on each plot outline the 5\% and 0.1\% of the peak emission. The analysis was repeated for the peak flux ratio as well. FCF$_{\mbox {\tiny peak}}$ results shows larger deviation from the mean, particularly at 450$\mu$m. This is to be expected given it is more susceptible to focus and temperature variations in the optics, however, again, no systematic deviation is observed. The resulting FCFs at each wavelength are thus:

\begin{equation}
{\mbox {FCF[850]}}_{\mbox {\tiny arcsec}} = 2.34 \pm 0.08 \;\;{\mbox {Jy/pW/arcsec}}^2
\end{equation}

and
\begin{equation}
{\mbox {FCF[450]}}_{\mbox {\tiny arcsec}} = 4.71 \pm 0.5  \;\;{\mbox {Jy/pW/arcsec}}^2
\end{equation}

The corresponding peak FCF results are:
\begin{equation}
{\mbox {FCF[850]}}_{\mbox {\tiny peak}} = 537 \pm 24 \;\;{\mbox {Jy/pW}}
\end{equation}

and
\begin{equation}
{\mbox {FCF[450]}}_{\mbox {\tiny peak}} = 491 \pm 67  \;\;{\mbox {Jy/pW}}
\end{equation}

\section{Observing modes} 

   \begin{figure}
   \begin{center}
  \begin{tabular}{c}
\includegraphics[height=12cm]{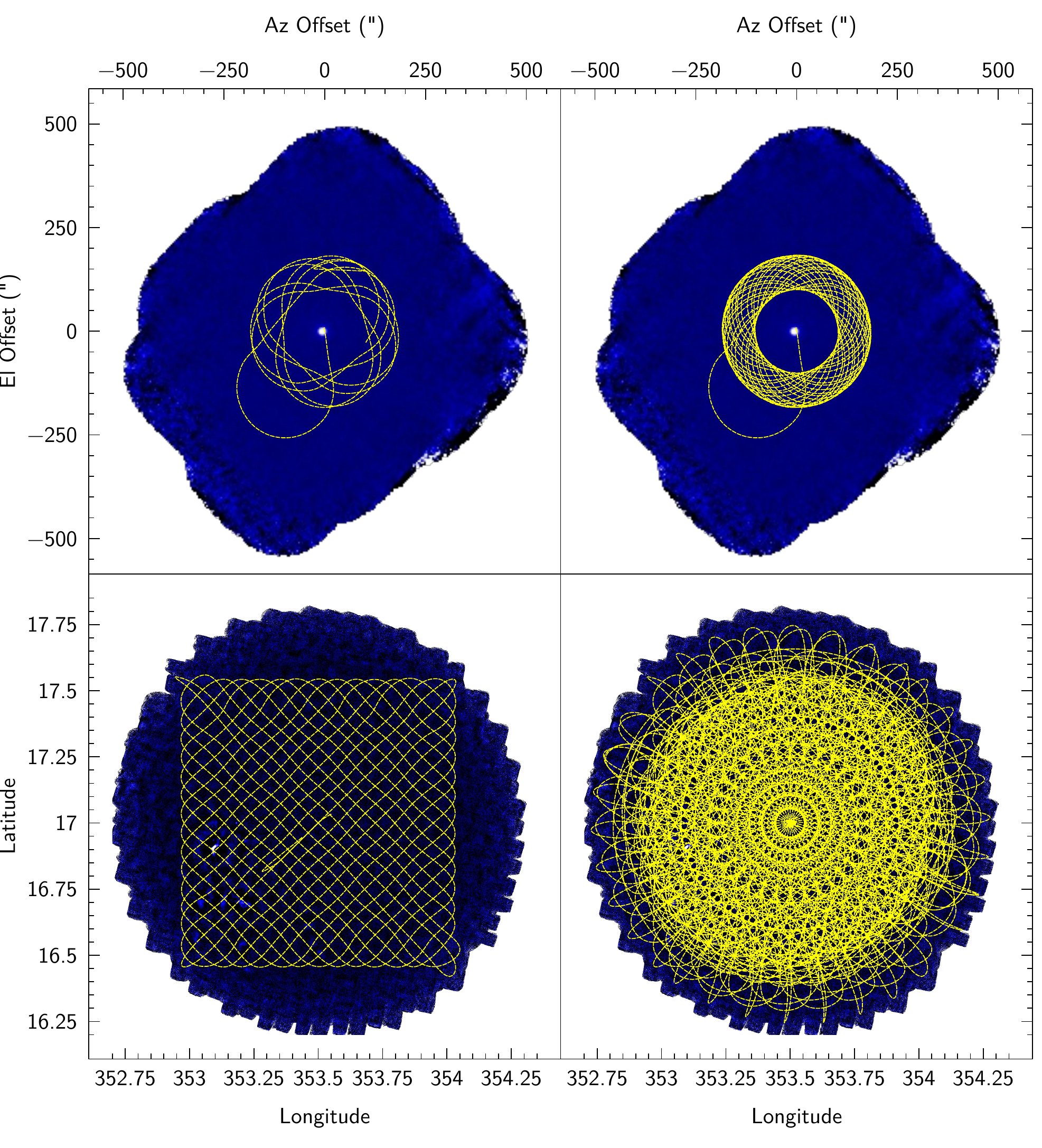}
  \end{tabular}
   \end{center}
   \caption[]{Plots of the telescope tracking position for a ``daisy'' pattern (top) and an 1800 arcsec diameter ``pong'' (bottom). The top left plot shows the mapping daisy pattern after 30 seconds integration while the top right shows the trace of the entire 4 minute observation. The bottom left image shows a single completed ``pong'' scan of the requested 1800 $\times$ 1800 arcsec area. This pattern is then repeated several times at equidistant angles to give the full map coverage shown on the right (the integration typically lasting 40 minutes). Note that this results in a circular field. \label{fig:patterns} }
   \end{figure} 
 
   \begin{figure}
   \begin{center}
   \begin{tabular}{c}
   \includegraphics[height=7cm]{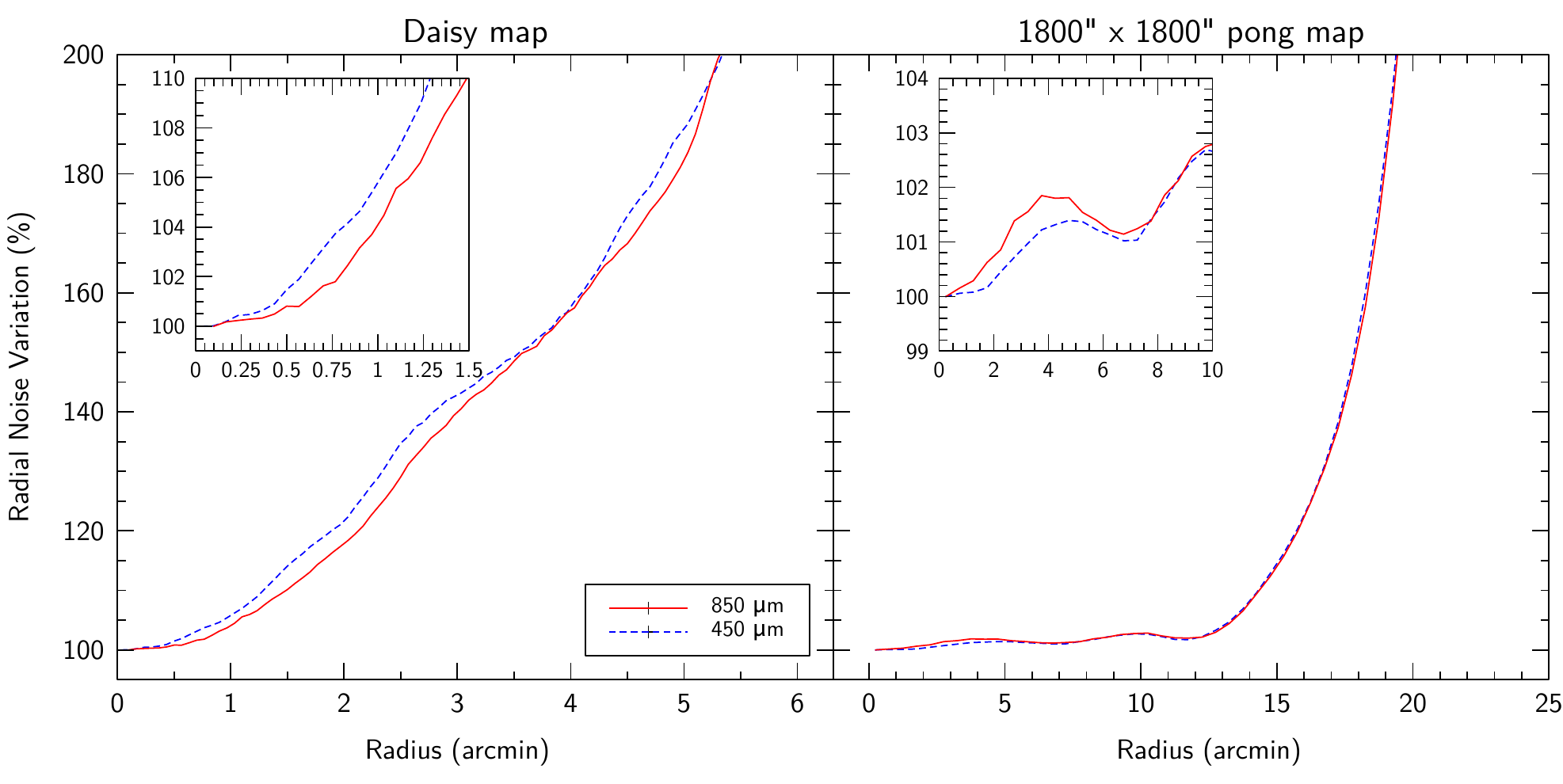}
   \end{tabular}
   \end{center}
   \caption{Radial noise proﬁprofile for a daisy scan (left) and 1800$\times$1800 pong scan (right). The percentage increase in the rms noise is plotted as a function of map radius.
 \label{fig:exptime} 
}
   \end{figure} 
 
All SCUBA-2 observations are conducted in scanning mode, with the scan speed high enough to place the signal from a point source at a frequency above the dominant 1/f noise in the bolometer spectrum. An optimum scanning mode should first and foremost mitigate the effects of 1/f drift, which might be achieved with a fast raster pattern, for example. However, if a scan strategy is too regular, any other periodic noise features become interlaced with the repeating pattern, producing scan-synchronous noise, which is extremely difficult to separate from the signal. Therefore, randomising a scan, or creating scan strategies that visit each region of the sky as many different time scales, and scan angles as possible offer the best protection against large-scale noise.\\

The constant speed ``daisy'' scans are the preferred observing pattern for fields smaller than the focal-plane ($<$8 arcmin). This pattern, shown in Figure~\ref{fig:patterns}, keeps the target coordinate on the array throughout the integration. The pattern size and turning radius were optimised to minimise the noise, maximise the on-source integration time, and produce the most uniform coverage. A typical daisy scan is shown in Figure~\ref{fig:patterns}, with a scan speed of 155 arcsec sec$^{-1}$ and a daisy size and turning radius equal to one quarter of the array size. The daisy pattern has the primary issue of failing at elevations $>$70$^\circ$ where it exceeds the telescope acceleration limits.\\

For larger scale mapping, the optimal scanning technique has been labeled the ``pong'' pattern. The bottom left plots in Figure~\ref{fig:patterns} show a single scan track of the telescope for an 1800$\times$1800 arcsec pong. Once this track is completed, the pattern is rotated and repeated at a new angle to improve the spatial modulation, until full coverage is achieved as shown in the bottom right hand plot in Figure~\ref{fig:patterns}. The telescope speed, pattern spacing, rotation number and rotation angles have been selected to produce the most uniform exposure time coverage over the widest possible field coverage. Currently the pong pattern parameters have been optimised for 900, 1800, 3600 and 7200 arcsec diameter fields.\\

The radial noise variation as a function of radius from the centre, for a typical daisy and 1800$\times$1800 pong, are shown in Figure~\ref{fig:exptime}. The left-hand plot shows the smoothly-varying noise in the daisy, which is optimised for peak exposure at the centre of the map, in contrast to the nearly-flat noise response out to 15 arcmin radius (the defined observing area) for the 1800$\times$1800 pong. \\

\section{Data reduction} 
 \begin{figure}
   \begin{center}
   \begin{tabular}{c}
   \includegraphics[height=8cm]{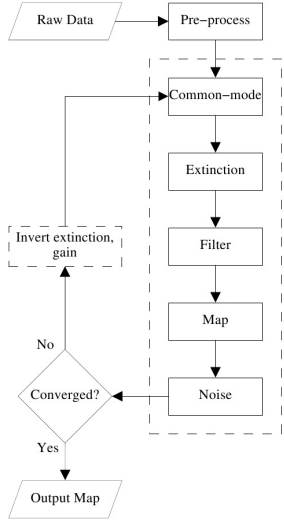}
   \end{tabular}
   \end{center}
   \caption{ Figure 17. The SMURF map-making algorithm presented as a flowchart showing the various model components in the iterative process \cite{chapin}. \label{fig:datared} 
}
   \end{figure} 
 
The SMURF map-maker is designed to achieve close to publication-level results in a reduction period close to the time taken to complete an observation, when using a high-end multi-core computer. This design goal was set to allow real-time feedback to observers at the telescope as well as providing a repository of acceptable quality data products. SCUBA-2 produces nearly unprecedented data volumes and also needs a more flexible data reduction technique than more tailored, memory-intensive algorithms such as CMB-type maximum likelihood map-makers. For these reasons, the basic method of the SMURF map-maker is to iteratively model and then remove the dominant correlated noise shared by all bolometers, and then remove the independent low-frequency noise, before reconstructing the residual signal. A detailed description of the map-maker is given in Ref.~\citenum{chapin} and \citenum{jenness2011}.\\

A flow-chart of the basic map-making algorithm is shown in Figure~\ref{fig:datared}. The full time-series of raw data is passed through pre-processing where flat-fielding removes bad bolometers (see Section~\ref{flat}) and spurious steps in the signal are removed. The common-mode signal, or correlated noise across the arrays, is then estimated by averaging the signal from all working bolometers. This common-mode signal is usually dominated by sky variation. The time-dependent atmospheric attenuation is then applied  from the WVM measurements (Section~\ref{atm}). The residual, and usually low-frequency, noise is then removed using a high-pass filter. The remaining corrected and cleaned data is then regridded using nearest-neighbour sampling. Each map pixel is now the average of many different bolometer samples, and is thus much less noisy than the original time-series. This initial map allows an estimation of the model astronomical signal which is removed to produce a residual time-series, which allows calculation of the white noise level in each detector. Each signal component can be influenced slightly by other component products, and so the extinction correction and gains from the common-mode are removed and the process is iterated until a specified noise level, or map tolerance, is attained. The configuration parameters for the map-maker are extremely versatile and nearly all component parameters can be user-defined. Standard configuration parameter files exist, and are still being refined, for reducing point-source, large-extended structures and blank/faint fields, for example.\\

\section{NEFDs} 
 \begin{figure}
   \begin{center}
   \begin{tabular}{c}
   \includegraphics[height=8cm]{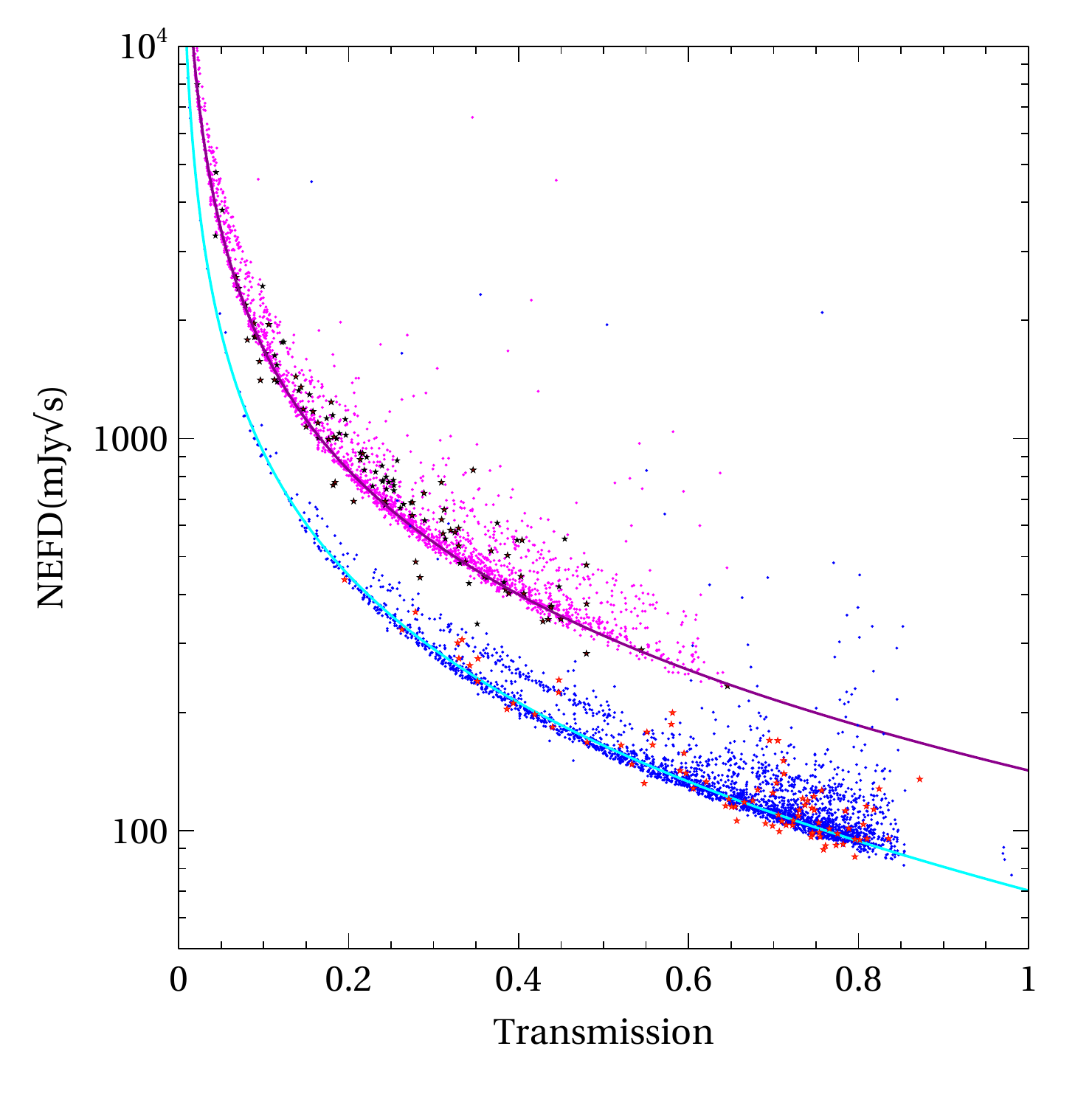}
   \end{tabular}
   \end{center}
   \caption{NEFDs for the 450$\mu$m (blue) and 850$\mu$m wavebands as derived from the measurements of the sky NEPs shown in Figure~\ref{fig:nepboth}. The fitted empirical relations with respect to the transmission,${\mbox{T}}_{\lambda}$,  are given by: ${\mbox {NEFD}}_{850} = [93.5/{\mbox{T}}_{850} - 23.9]$ mJy$\sqrt{s}$ at 850$\mu$m (cyan line) and  ${\mbox {NEFD}}_{450} = [171.9/{\mbox{T}}_{450} - 29.5]$ mJy$\sqrt{s}$ at 450$\mu$m (red line). NEFDs calculated from the variance and exposure time for a sample of calibration maps are plotted at 850/450$\mu$m (red/black star points) showing that the NEP-estimated NEFDs agree well, though they perhaps slightly under-estimate the sensitivity achieved by the astronomical maps, likely a result of the improved bolometer-flagging and other optimisations in the reduction software.\label{fig:nefd} 
}
   \end{figure} 

 \begin{figure}
   \begin{center}
   \begin{tabular}{c}
   \includegraphics[height=6cm]{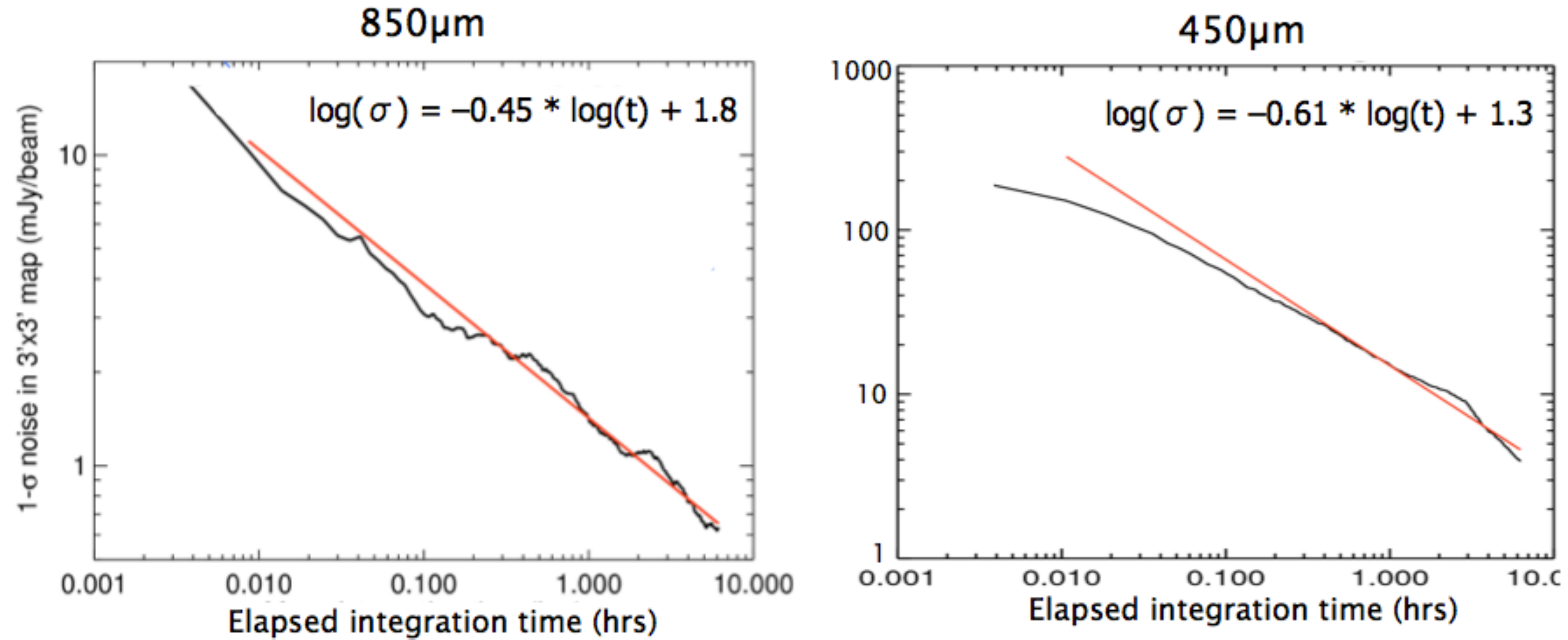}
   \end{tabular}
   \end{center}
   \caption{Measured 1$\sigma$ noise in a 3 arcmin diameter daisy image as a function of observing time for a 7 hr observation
taken in good observing conditions ($\tau_{\mbox {\tiny 225}} \sim$ 0.05) for the 850$\mu$m (left) and 450$\mu$m (right) wavebands. The red line is a fit to the data as given by the equations in each panel.\label{fig:integ} 
}
   \end{figure}

\begin{table}[h]
\caption{Observing time required to reach target rms levels with SCUBA-2}
\label{etime}
\begin{center}
\begin{tabular}{|c|c|c|c|c|} 
\hline
\rule[-1ex]{0pt}{3.5ex}  Wavelength & Obs mode & Target rms & Transmission  & Observing time \\
\rule[-1ex]{0pt}{3.5ex}  [$\mu$m] & --   & [mJy] & \%  &  [hrs]\\
\hline
\rule[-1ex]{0pt}{3.5ex}   850 & Daisy  & 1.0  & 80   &  2.5 \\
\hline
\rule[-1ex]{0pt}{3.5ex}   850 & Daisy  & 5.0  & 80   &  0.1 \\
\hline
\rule[-1ex]{0pt}{3.5ex}   850 & Pong 3600$\times$3600 & 5.0  & 80  &  7.5 \\
\hline
\rule[-1ex]{0pt}{3.5ex}   850 & Pong 3600$\times$3600  & 10.0  & 80   &  2.0 \\
\hline
\rule[-1ex]{0pt}{3.5ex}   450 & Daisy  & 5.0  & 50   & 6.8 \\
\hline
\rule[-1ex]{0pt}{3.5ex}   450 & Daisy  & 30.0  & 50  &  0.2 \\
\hline
\rule[-1ex]{0pt}{3.5ex}   450 & Pong 3600$\times$3600 & 30.0  & 50  & 21.0 \\
\hline
\rule[-1ex]{0pt}{3.5ex}   450 & Pong 3600$\times$3600  & 60.0  & 50  &  5.0 \\
\hline
\end{tabular}
\end{center}
\end{table}
The noise equivalent flux density (NEFD) is defined as the astronomical source strength required to reach a S/N ratio of one in a second of integration time. In practice, this defines the sensitivity limits of the instrument and can be used to predict the integration time required to reach a desired rms noise level. The NEFDs for SCUBA-2 have been derived using the measured sky NEPs shown in Section~\ref{nep}, as follows for each waveband: ${\mbox {NEFD}}_{(\lambda)} =  ({\mbox {NEP}}_{\mbox{\tiny sky}} \cdot {\mbox {FCF}}_{(\lambda)})/{{\mbox{T}}_{(\lambda)}}$ and is given in units of ${\mbox {mJy}}\sqrt{s}$ and the transmission ${\mbox{T}}_{(\lambda)}$ is calculated using the line-of-sight extinction $\tau_{\mbox {\tiny los}}$ as given by relations in Section~\ref{atm} by ${\mbox{T}}_{(\lambda)} = e^{(-\tau_{\mbox {\tiny los}})}$. The average per-bolometer NEFDs are shown in Figure~\ref{fig:nefd} for both wavebands. In addition, the plot shows NEFDs calculated from the noise and exposure time in a sample of calibration maps, which show good agreement with the NEP-derived values. The fits to these results can be used to calculate the integration time required to reach a required sensitivity limit at each wavelength. Figure~\ref{fig:integ} shows the measured noise in a map as a function of integration time, and the fit shows that the noise decreases as expected as a function of $\sqrt{t}$. Table~\ref{etime} uses the NEFD relations in Figure~\ref{fig:nefd} to calculate the observing time required to reach target rms noise levels for different observing modes.

\section{Science on sky} 
  \begin{figure}
   \begin{center}
   \begin{tabular}{c}
   \includegraphics[height=4cm]{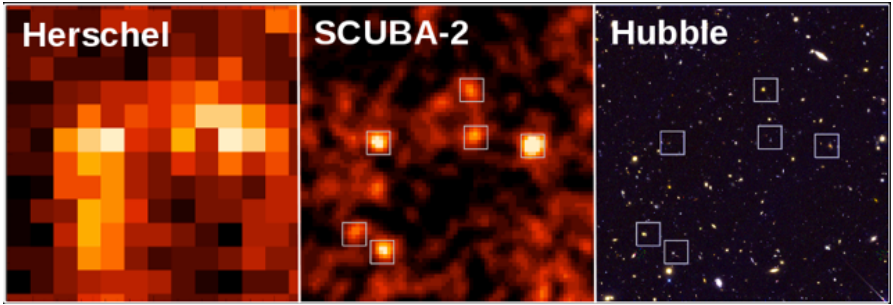}
   \end{tabular}
   \end{center}
   \caption{450$\mu$m image of a deep cosmology field showing six distinct submillimetre sources, which were blended together in the poorer-quality Herschel image (shown on the left). These six SCUBA-2 sources can then be matched to galaxies in the Hubble Space Telescope image (on the right). Credit: ROE/JCMT/SCUBA-2 image courtesy of SC2CLS survey team.\label{cls}}
   \end{figure} 

 \begin{figure}
   \begin{center}
   \begin{tabular}{c}
   \includegraphics[height=7cm]{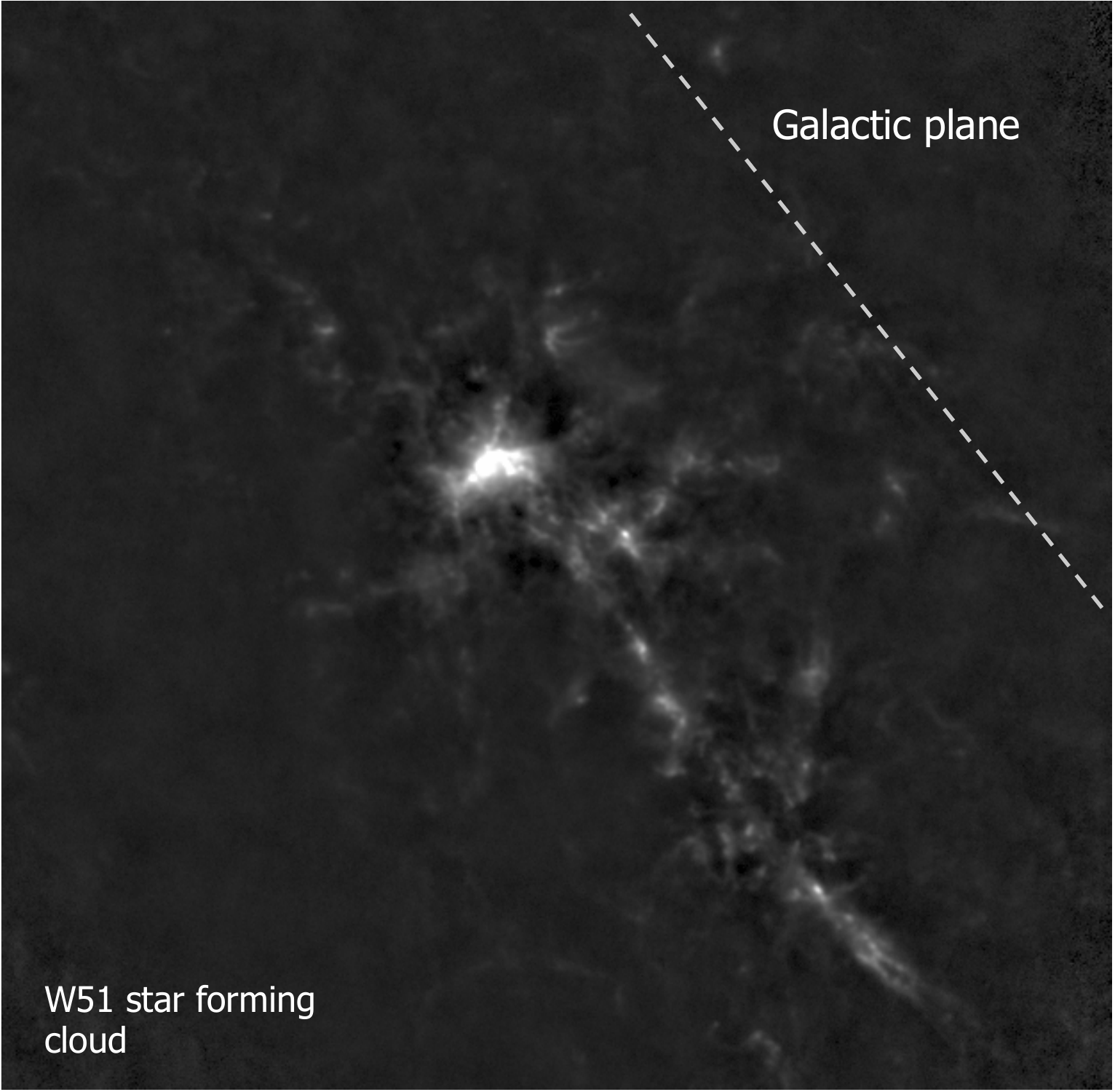}
   \end{tabular}
   \end{center}
   \caption{A 1$\times$1 degree image of high-mass star forming region W51 in Aquila as observed by SCUBA-2 at 850$\mu$m. Courtesy JPS team. \label{w51} 
}
   \end{figure} 

 \begin{figure}
   \begin{center}
   \begin{tabular}{c}
   \includegraphics[height=7cm]{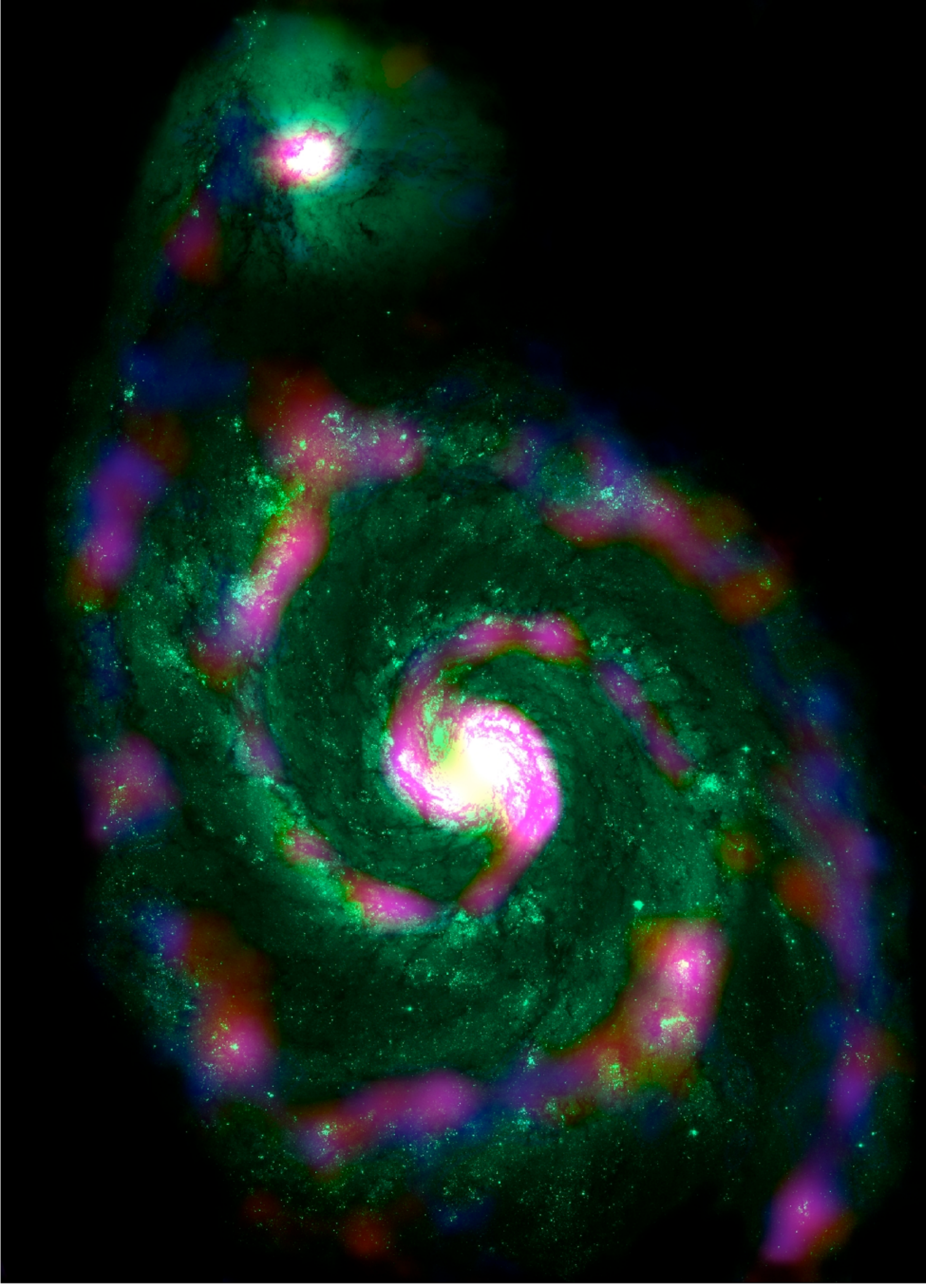}
   \end{tabular}
   \end{center}
   \caption{A composite image of M51, the Whirlpool Galaxy, with the 450$\mu$m (red) and 850$\mu$m (blue) emission detected by SCUBA-2 superimposed on a greyscale HST image. Credit: JCMT, UBC, NASA/ESA/S. Beckwith (STScI) \& Hubble Heritage Team. \label{m51} 
}
   \end{figure} 

SCUBA-2 was released to the community on October 2011. The first science users were from the JCMT Legacy Surveys (JLS) and the first PI observers came to the telescope in February 2012.In this relatively short time, SCUBA-2 has already produced some exciting science results, with over 1000 hours already observed on the sky since February 2012 (at the time of writing). The stand-out capabilities of SCUBA-2, a wide field-of-view and fast mapping speed, lends itself to sensitive surveys of the submillimetre sky, and the JLS projects are designed with science goals which take advantage of this.\\

Already, the early science from the instrument is providing exciting glimpses to the full impact and legacy that SCUBA-2 will ultimately have. For the first time, the population of galaxies that are thought to make up the Cosmic Infrared Background have been resolved at 450$\mu$m. Figure~\ref{cls} shows a section of a deep map from the SCUBA-2 Cosmology Legacy Survey showing a comparison of SCUBA-2 at 450$\mu$m, Herschel/SPIRE at 500$\mu$m and HST. The superior resolution of SCUBA-2 compared to Herschel/SPIRE is clear to see and, for the first time, submillimetre galaxies detected at 450$\mu$m are being directly identified by the optical counterparts. Figure~\ref{w51} shows a  1-square degree map of the W51 high-mass star forming region in Aquila, taken at 850$\mu$m. SCUBA-2 easily detects the ridge of cold gas running parallel to the Galactic Plane. Studies of this kind will probe the high-mass end of the stellar initial mass function, and the sensitivity of SCUBA-2 is adequate to detect all high-mass and cluster star-formation in our galaxy. Figure~\ref{m51} shows the greyscale HST image of the Whirlpool Galaxy, M51,  overlaid with the SCUBA-2 450$\mu$m (red) and 850$\mu$m (blue) emission. The nuclei of the two interacting galaxies are detected by SCUBA-2 at both wavelengths. The spatial resolution reveals details of regions of hot star formation in the outer arms of the galaxy.\\

\section{Conclusions} 

SCUBA-2 has been commissioned and operational for science observations since October 2011. The phases of on-sky commissioning were discussed, including the flat-fielding methodology and absolute-power calibration. The noise performance of the instrument was presented for the first science semester, beginning in February 2012. The NEPs show good stability over both short and long time scales, though some unknown noise features are observed and are under investigation. The atmospheric calibration technique was outlined and the waveband-dependent extinction correction terms were shown. The flux calibration has been successfully characterised, with absolute deviation in the FCF performance of less than 10\% at both wavelengths. Description of the observing modes and data-reduction were given. Finally, the instrument sensitivities were calculated, showing that in the best weather, a 1-square-degree observation reaches 10 mJy sensitivity in approximately 2hrs at 850$\mu$m, while at 450$\mu$m, the same size map can reach 60 mJy rms in 5 hours.\\ 

\section{Acknowledgments}
The SCUBA-2 project is funded by the UK Science and Technology Facilities Council (STFC), the JCMT Development Fund and the Canadian Foundation for Innovation (CFI).


\bibliography{scuba2_spie2012}  
\bibliographystyle{spiebib}  
\end{document}